\documentclass[%
 reprint,
 amsmath,amssymb,
 aps,
floatfix,
]{revtex4-2}

\usepackage{graphicx}
\usepackage{dcolumn}
\usepackage{bm}

\usepackage{braket}
\usepackage{color}

\begin{document}

\preprint{APS/123-QED}

\title{Exploring exact-factorization-based trajectories for low-energy dynamics near a conical intersection}

\author{Lea M. Ibele}
\email{lea-maria.ibele@universite-paris-saclay.fr}
\affiliation{Universit\'e Paris-Saclay, CNRS, Institut de Chimie Physique UMR8000, 91405, Orsay, France}%
\author{Federica Agostini}
\affiliation{Universit\'e Paris-Saclay, CNRS, Institut de Chimie Physique UMR8000, 91405, Orsay, France}%
\date{\today}

\onecolumngrid 
\begin{abstract} 
We study low-energy dynamics generated by a two-dimensional two-state Jahn-Teller Hamiltonian in the vicinity of a conical intersection using quantum wavepacket and trajectories dynamics. Recently, these dynamics were studied by comparing the adiabatic representation and the exact factorization, with the purpose to highlight the different nature of topological- and geometric-phase effects arising in the two theoretical representation of the same problem. Here, we employ the exact factorization to understand how to model accurately low-energy dynamics in the vicinity of a conical intersection using an approximate description of the nuclear motion that uses trajectories. We find that, since nonadiabatic effects are weak but non-negligible, the trajectory-based description that invokes the classical approximation struggles to capture the correct behavior.
\end{abstract}

\maketitle

\onecolumngrid 
\section{Introduction}
Conical intersections, points of degeneracy between two adiabatic potential energy surfaces are ubiquituous in molecular physics and photochemistry. Due to the strong nonadiabatic electron-nuclear coupling in their vicinity, they act as funnels for electronically excited population to decay to its ground state and therefore strongly influence the photoinduced processes in molecules.\cite{Baer_book2006,yarkony1996diabolical,domcke2004conical,boeije2023one}
If those degeneracies occur between the excited state and the ground state, they can also influence low-energy processes and induce a nonadiabatic behavior of the dynamics of a system evolving mainly in the ground state.\cite{ryabinkin2013geometric,kendrick1996geometric,bernardi1990mechanism,celani1997conical} In the past, such influences have been postulated and discussed for transition states where higher lying electronically excited states are involved.\cite{desouter1983transition,carpenter2022conical}
As conical intersections represent points of degeneracies between adiabatic states and, thus, of a singular nonadiabatic coupling, it has been postulated that they can induce a \textit{molecular geometric phase} on the dynamics.\cite{mead1980molecular,hazra2016geometric,kendrick2015geometric,ribeiro2018vibronic,juanes-marcos2005theoretical,yuan2018observation,yuan2020observation,yuan2018direct,xie2020quantum,daoud2018exploring}
This means that a nuclear wavepacket picks up a phase when encircling a conical intersection, and if parts of the nuclear wavepacket pass the conical intersection on different sides, they acquire opposite phases leading to destructive interference upon recombination.

This discussion intrinsically assumes an adiabatic representation of the photochemical or low-energy process. 
In this representation, the molecular wavefunction is expressed in a basis of adiabatic electronic states, obtained from the electronic Hamiltonian, with the corresponding nuclear coefficients.
Since the electronic Hamiltonian depends on the nuclear positions, the adiabatic states and the corresponding eigenvalues, i.e. the adiabatic potential energy surfaces, aquire a characteristic dependence on the nuclear positions. This dependence produces degeneracies, singularities and, thus, conical intersections, which become signatures of the adiabatic representation.
 
Recently, an experiment mimicking a nuclear wavepacket encircling a conical intersection has been performed for a two-dimensional Jahn-Teller Hamiltonian using a mixed qudit-boson simulator.\cite{valahu2023direct} Destructive interference between the wavepacket parts has been reported  leading to a nodal-like feature in the nuclear density which has been proposed as a direct experimental observation of the geometric phase.
We studied\cite{ibele2023nature} the same system through numerical quantum dynamics simulations and analyzed the dynamics comparing the adiabatic representation with the exact factorization.
Our study showed that in the adiabatic representation the geometric phase is actually topological, as it is well-known in the presence of conical intersections~\cite{truhlar1980determination, mead1980molecular}. However, such a topological feature is not robust under changes of representation as it disappears when the exact factorization is employed for the analysis. A physical observable, coined \textit{dynamics-induced geometric phase}, can be introduced instead, independent of the representation which is, therefore, unrelated to conical intersections.\cite{ibele2023nature,martinazzo2023dynamics,requist2016molecular}

The exact factorization is a representation of the molecular wavefunction that writes it as a single product of time-dependent electronic and nuclear wavefunction. 
This formalism\cite{abedi2010exact,abedi2012correlated,Gross_EPJB2021,Curchod_WIRES2019} gives rise to single, time-dependent scalar and vector potentials governing the nuclear wavefunction evolution. Thus, it has been shown in the past that these potentials allow one to introduce in a natural way the concept of classical-like force to drive the nuclear dynamics and, thus, are well-suited for the development of trajectory-based algorithms.\cite{agostini2018nuclear,talotta2020quantum,min2017ab,ibele2022photochemical} In this work, our aim is to investigate the time-dependent scalar and vector potentials of the low-energy dynamics around a conical intersection of Refs.~[\citenum{valahu2023direct, ibele2023nature}], in relation to various trajectory schemes based on the exact factorization. Such an analysis will allow us to understand if and how one can circumvent including the geometric phase of the adiabatic representation when an approximate description of the dynamics using nuclear trajectories is employed.

In this paper, Section~\ref{sec: methods} is devoted to the presentation of the exact factorization and of the trajectory schemes that will be used for the analysis of the dynamics. Section~\ref{sec: comp details} introduces the model Hamiltonian and provides information on the simulations. In Section~\ref{sec: results}, we report our results and our interpretation of the trajectories dynamics. Using the time-dependent potentials obtained from the numerically exact quantum dynamics, we first look at quantum trajectories, then we study the influence of the classical approximation on the trajectories' dynamics, in order to understand the effect of the different terms of the time-dependent potentials. Based on this analysis, we investigate why the coupled-trajectory mixed-quantum-classical (CT-MQC) algorithm\cite{min2015coupled,min2017ab,curchod2018ct} fails to reproduce the correct dynamics and we compare it to adiabatic-like dynamics using, as done previously, quantum and classical trajectories. We conclude in Section~\ref{sec: concl} by summarizing the shortcomings of the approximate CT-MQC scheme for such a dynamical problem.

\section{Methods}\label{sec: methods}
The time evolution of the electron-nuclear wavefunction describing the state of a molecule, $\Psi(\boldsymbol r, \boldsymbol R, t)$ is governed by the time-dependent Schr\"odinger equation
\begin{equation}\label{eqn: TDSE}
i\hbar \frac{\partial}{\partial t}\Psi(\boldsymbol r, \boldsymbol R, t)=\hat{H}(\boldsymbol r,\boldsymbol R)\Psi(\boldsymbol r, \boldsymbol R, t)
\end{equation}
with the full, molecular Hamiltonian $\hat{H}(\boldsymbol r, \boldsymbol R)$ consisting of the sum of the nuclear kinetic energy operator $\hat{T}_{\text{n}}(\boldsymbol R)$ and the electronic, Born-Oppenheimer Hamiltonian $\hat{H}_{\text{BO}}(\boldsymbol r,\boldsymbol R)$, consisting of the electronic kinetic energy and of the interaction potential. $\boldsymbol r$ and $\boldsymbol R$ are collective variables for the electronic and nuclear coordinates of the molecule, respectively.

\subsection{Exact Factorization of the Molecular Wavefunction}
The exact factorization\cite{abedi2010exact,abedi2012correlated,Gross_EPJB2021, Curchod_WIRES2019} proposes to express the molecular wavefunction as a single product of time-dependent nuclear $\chi(\boldsymbol R,t)$ and electronic $\Phi_{\boldsymbol{R}}(\boldsymbol r,t)$ wavefunctions
\begin{equation}
\Psi(\boldsymbol r, \boldsymbol R, t)=\Phi_{\boldsymbol{R}}(\boldsymbol r,t)\chi(\boldsymbol R,t)
\label{eq:EF}
\end{equation}
It is important to note that the electronic wavefunction depends parametrically on the nuclear coordinates, denoted by the subscript $\boldsymbol R$. Once the partial normalization condition is imposed, i.e. $\braket{\Phi_{\boldsymbol{R}}(t)|\Phi_{\boldsymbol{R}}(t)}_{\boldsymbol r} = 1 \forall \boldsymbol R, t$ (the notation $\braket{.}_{\boldsymbol r}$ signifies integration over the whole space of electronic coordinates), it can be proven that the single product form~(\ref{eq:EF}) is exact and unique up to a gauge-like transformation.\cite{abedi2010exact,Curchod_WIRES2019} In the current work we fix the gauge so that the nuclear wavefunction is real and non-negative, i.e. $\chi(\boldsymbol R,t)=|\chi(\boldsymbol R,t)| \forall \boldsymbol R, t$. 

The time-evolution of the molecular wavefunction can be obtained within the exact factorization upon inserting the factored form~(\ref{eq:EF}) into the time-dependent Schr\"odinger equation~(\ref{eqn: TDSE}). Using the partial normalization condition, we can derive coupled equations of motion for the nuclear and electronic wavefunctions:
\begin{align}
i\hbar \frac{\partial}{\partial t}\chi(\boldsymbol R,t)&=\left[\sum_\mu^{N_n} \frac{[-i\hbar \nabla_{\mu} + \boldsymbol A_\mu(\boldsymbol R,t)]^2}{2M_\mu} + \epsilon(\boldsymbol R,t) \right]\chi(\boldsymbol R,t)  \label{eq:eom_nu}\\
i\hbar \frac{\partial}{\partial t}\Phi_{\boldsymbol R}(\boldsymbol r,t)&=\left[\hat{H}_{\text{BO}}+\hat{U}_{\text{en}}[\Phi,\chi](\boldsymbol R,t)-\epsilon(\boldsymbol R,t) \right]\Phi_{\boldsymbol R}(\boldsymbol r,t)
\label{eq:eom_el}
\end{align}
where $\mu$ denotes the $N_n$ nuclei. The evolution equation for the nuclear wavefunction couples to the electronic wavefunction implicitly through the two new potentials, the time-dependent vector potential (TDVP) $\boldsymbol A_\mu(\boldsymbol R,t)$,
\begin{equation}\label{eq:TDVP}
\boldsymbol A_\mu(\boldsymbol R,t)=\braket{\Phi_{\boldsymbol R}(t)|-i\hbar \nabla_\mu\Phi_{\boldsymbol R}(t)}_{\boldsymbol r}
\end{equation}
and the time-dependent potential energy surface (TDPES) $\epsilon(\boldsymbol R,t)$
\begin{equation}
\epsilon(\boldsymbol R,t)=\bra{\Phi_{\boldsymbol R}(t)}\hat{H}_{\text{BO}}+\hat{U}_{\text{en}}[\Phi,\chi](\boldsymbol R,t)-i\hbar\frac{\partial}{\partial t}\ket{\Phi_{\boldsymbol R}(t)}_{\boldsymbol r}
\end{equation}
The electronic evolution equation is coupled to the nuclear wavefunction through the electron-nuclear coupling operator, $\hat{U}_{\text{en}}[\Phi,\chi](\boldsymbol R,t)$, which explicitly depends on the nuclear wavefunction and also implicitly contains the electronic wavefunction through the TDVP: 
\begin{equation}
\hat{U}_{\text{en}}[\Phi,\chi](\boldsymbol R,t)=\sum_\mu^{N_n}\frac{1}{M_\mu}\left[\frac{[-i\hbar\nabla_\mu - \boldsymbol A_\mu]^2}{2}+\left(\frac{-i\hbar\nabla_\mu\chi}{\chi}+\boldsymbol A_\mu \right)\left( -i\hbar \nabla_\mu - \boldsymbol A_\mu \right)  \right]
\label{eq:Uen}
\end{equation}
Note that for conciseness the $(\boldsymbol R,t)$ dependencies of $\chi$ and $\boldsymbol A_\mu$ have been dropped.

As introduced above, the exact factorization of the molecular wavefunction is unique up to a gauge transformation, 
\begin{equation}
\Psi(\boldsymbol r, \boldsymbol R, t)=\Phi_{\boldsymbol{R}}(\boldsymbol r,t)\chi(\boldsymbol R,t)
=\Phi_{\boldsymbol{R}}(\boldsymbol r,t)e^{-i/\hbar\theta(\boldsymbol R,t)}\chi(\boldsymbol R,t)e^{i/\hbar\theta(\boldsymbol R,t)}=\tilde{\Phi}_{\boldsymbol{R}}(\boldsymbol r,t)\tilde{\chi}(\boldsymbol R,t)
\end{equation}
for a real gauge function $\theta(\boldsymbol R,t)$. While the nuclear and electronic equations of motion, Eqs.~(\ref{eq:eom_nu}) and (\ref{eq:eom_el}), respectively, are form-invariant under a gauge transformation, the TDVP and TDPES transform according to
\begin{align}
\tilde{\epsilon}(\boldsymbol R,t)=\epsilon(\boldsymbol R,t)+\frac{\partial}{\partial t}\theta(\boldsymbol R,t) \\
\tilde{\boldsymbol A}_\mu(\boldsymbol R,t)={\boldsymbol A}_\mu(\boldsymbol R,t)+\nabla_\mu\theta(\boldsymbol R,t)
\end{align}
The TDVP is inherently gauge dependent and the TDPES can be decomposed into two gauge-invariant ($\epsilon_{\text{BO}}$,$\epsilon_{\text{geo}}$) and one gauge-dependent contributions ($\epsilon_{\text{GD}}$):
\begin{align}
\epsilon_{\text{BO}} &=\bra{\Phi_{\boldsymbol R}(t)}\hat{H}_{\text{BO}}(\boldsymbol R)\ket{\Phi_{\boldsymbol R}(t)}_{\boldsymbol r} \label{eq:tdpes_bo}  \\
\epsilon_{\text{geo}} &= \sum_\mu \frac{\hbar^2\braket{\nabla_\mu \Phi_{\boldsymbol R}(t)|\nabla_\mu \Phi_{\boldsymbol R}(t)}_{\boldsymbol r}}{2M_\mu}-\sum_\mu \frac{|\boldsymbol A_\mu|^2}{2M_\mu} \label{eq:tdpes_geo} \\
\epsilon_{\text{GD}} &= \bra{\Phi_{\boldsymbol R}(t)}-i\hbar\frac{\partial}{\partial t}\ket{\Phi_{\boldsymbol R}(t)}_{\boldsymbol r} \label{eq:tdpes_gd}
\end{align}
The first of the two gauge-invariant contributions, $\epsilon_{\text{BO}}$, arises from the Born-Oppenheimer Hamiltonian and yields an mean-field-like contribution to the TDPES. The second term, $\epsilon_{\text{geo}}$, accounts for the spatial variation of the electronic wavefunction in the parametric space $\boldsymbol R$, as it is evident from its definition~(\ref{eq:tdpes_geo}). Therefore, we interpret it as somehow responsible for nonadiabatic effects in the TDPES. In addition, if an analogy is drawn between TDPES/TDVP and classical electromagnetism, it has been argued that the gradient of $\epsilon_{\text{geo}}$ gives rise to a geometric contribution to the electric-field-like effect of the electrons acting on the nuclei.\cite{requist2016molecular}

\subsection{Trajectory formalisms from the exact factorization}\label{sec:traj}
As ensured by the partial normalization condition, at all times $|\chi(\boldsymbol R,t)|^2$ reproduces the nuclear density obtained from $\Psi(\boldsymbol r, \boldsymbol R, t)$. Thus, we can equally express the nuclear wavefunction as
\begin{equation}
\chi(\boldsymbol R,t) = \sqrt{\braket{\Psi(\boldsymbol R,t)|\Psi(\boldsymbol R,t)}_{\boldsymbol r}} \exp\left(\frac{i}{\hbar}S(\boldsymbol R,t) \right)= |\chi(\boldsymbol R,t)| \exp\left(\frac{i}{\hbar}S(\boldsymbol R,t) \right)
\end{equation}
Inserting this polar form of the nuclear wavefunction into the time-dependent Schr\"odinger equation allows us to derive an evolution equation for the phase, as shown in Refs.~[\citenum{agostini2018nuclear,talotta2020quantum}]. This can be identified as a nuclear Hamilton-Jacobi equation using as canonical momentum $\nabla_\mu S(\boldsymbol R,t)=\tilde{\boldsymbol P}_\mu(\boldsymbol R,t)$:
\begin{equation}
-\frac{\partial}{\partial t}S(\boldsymbol R,t)=H_{\text{n}}(\tilde{\boldsymbol P},\boldsymbol R,t)=\sum_\mu\frac{[\tilde{\boldsymbol P}_\mu(\boldsymbol R,t)+\boldsymbol A_\mu(\boldsymbol R,t)]^2}{2M_\mu}+\epsilon(\boldsymbol R,t)+Q_{\text{pot}}(\boldsymbol R,t)
\label{eq:dSdt}
\end{equation}
The quantum potential, $Q_{\text{pot}}(\boldsymbol R,t)=
\sum_\mu\frac{-\hbar^2}{2M_\mu}\frac{\nabla_\mu^2|\chi(\boldsymbol R,t)|}{|\chi(\boldsymbol R,t)|}$, introduces most of the nuclear quantum effects, while the other contributing terms are the nuclear kinetic energy with the nuclear mechanical momentum field $\boldsymbol P_\mu(\boldsymbol R,t)=\tilde{\boldsymbol P}_\mu(\boldsymbol R,t)+\boldsymbol A_\mu(\boldsymbol R,t)$ and the ``classical'' potential energy $\epsilon(\boldsymbol R,t)$.

Using the method of characteristics to solve this Hamilton-Jacobi partial differential equations yields Hamilton-like evolution equations,
\begin{align}
\dot{\boldsymbol R}_\mu(t)&=\frac{\tilde{\boldsymbol P}_\mu(t)+\boldsymbol A_\mu(\boldsymbol R(t),t)}{M_\mu}  \\
\dot{\tilde{\boldsymbol P}}_\mu(t)&=-\nabla_\mu H_{\text{n}}(\tilde{\boldsymbol P}(t),\boldsymbol R(t),t)
\end{align}
where the canonical momentum appears, which are equivalent to
\begin{align}
\dot{\boldsymbol R}_\mu(t)&= \frac{\boldsymbol P_\mu(t)}{M_\mu} \\
\dot{\boldsymbol P}_\mu(t)&=-\nabla_\mu H_{\text{n}}(\boldsymbol P(t),\boldsymbol R(t),t)+\dot{\boldsymbol A}_\mu(\boldsymbol R(t),t)
\end{align}
using the mechanical momentum. Introducing the characteristics, i.e. a set of ordinary differential equations that always satisfy the original Hamilton-Jacobi equation, allows us to define the trajectories $\boldsymbol R(t)\equiv \{\boldsymbol R_\mu(t)\}_{\mu=1,\dots,N_n}$ and their corresponding momenta $\boldsymbol P(t)\equiv \{\boldsymbol P_\mu(t)\}_{\mu=1,\dots,N_n}$. 

In this work, we chose a gauge where the nuclear wavefunction is real and non-negative, thus, $S(\boldsymbol R,t)=0$ which yields the evolution equations 
\begin{align}
\dot{\boldsymbol R}_\mu(t)&=\frac{\boldsymbol A_\mu(\boldsymbol R(t),t)}{M_\mu}\label{eqn: R}  \\
\dot{\boldsymbol P}_\mu(t)&=\dot{\boldsymbol A}_\mu(\boldsymbol R(t),t)\label{eqn: TDVP as P}
\end{align}
Equation~(\ref{eqn: TDVP as P}) clearly shows that the nuclear momentum field is the TDVP. These evolution equations have been derived by simply introducing the characteristics to solve the nuclear Hamilton-Jacobi equation, Eq.~(\ref{eq:dSdt}), without invoking the classical approximation yet. Consequently, the ``trajectories'' emerging as solutions of the evolution equations~(\ref{eqn: TDVP as P}) and~(\ref{eqn: R}) are \textit{quantum} trajectories. For these quantum trajectories the TDVP is the only quantity driving the dyanmics.\cite{ibele2022photochemical}

Once we apply the classical approximation, i.e. neglecting the quantum potential in Eq.~(\ref{eq:dSdt}), the evolution equations become
\begin{align}
\dot{\boldsymbol R}_\mu(t)&= \frac{\boldsymbol P_\mu(t)}{M_\mu} \\
\dot{\boldsymbol P}_\mu(t)&=-\nabla_\mu H^{\text{cl}}_{\text{n}}(\boldsymbol P(t),\boldsymbol R(t),t)+\dot{\boldsymbol A}_\mu(\boldsymbol R(t),t)
\end{align}
with classical Hamiltonian
\begin{equation}
H^{\text{cl}}_{\text{n}}(\boldsymbol P(t),\boldsymbol R(t),t)=\sum_\mu \frac{\boldsymbol P_\mu(\boldsymbol R,t)^2}{2M_\mu}+\epsilon(\boldsymbol R,t)
\end{equation}
Having found evolution equations for both quantum and classical trajectories, we only need to discuss how the initial conditions are selected. For this, it is important to note that for the quantum trajectories, position and momenta are not independent variables. Therefore, the initial positions, for both quantum and classical trajectories, can be sampled stochastically from the initial nuclear probability distribution. While for the classical trajectories the momenta can equally be stochastically and independently sampled from a momentum probability distribution, e.g. calculated from the Wigner transform of the initial nuclear wavefunction, for the quantum trajectories the initial momenta need to be selected as $\boldsymbol A_\mu(\boldsymbol R_0,t=0)$ for an ensemble of initial positions $\boldsymbol R_0$.

Analyzing the nuclear dynamics using these quantum and classical trajectories allows us to focus purely on the accuracy of trajectories in capturing the simulated dynamics. The potential driving the dynamics are numerically exact, as they are determined from the solution of the full time-dependent Schr\"odinger equation. The exact factorization is a suitable formalism for such an analysis because it is not based on the adiabatic representation and inherently couples the nuclear dynamics on different electronic states. 

\subsection{Coupled-trajectory mixed quantum-classical approach}\label{sec:ctmqc}
Having introduced trajectory formalisms to propagate nuclear trajectories using the exact TDVP and TDPES, we briefly introduce in this section a formalism that approximates the TDVP and TDPES from the electronic wavefunction, which allows one to simulate nonadiabatic dynamics on-the-fly: the so-called Coupled-Trajectory Mixed Quantum-Classical (CT-MQC) algorithm.
We outline here just the main ideas of this formalism and refer the interested reader to more in depth descriptions in literature.\cite{min2017ab,talotta2020internal}

CT-MQC has been derived~\cite{min2015coupled} starting from the evolution equation of the phase of the nuclear wavefunction (Eq.~(\ref{eq:dSdt})) following the procedure detailed in Section~\ref{sec:traj} to derive Eqs.~(\ref{eqn: R}) and~(\ref{eqn: TDVP as P}). However, in CT-MQC the gauge is fixed so that $\epsilon(\boldsymbol R(t),t) + \sum_{\mu'}\dot{\boldsymbol R}_{\mu'}(t)\cdot \boldsymbol A_{\mu'}(\boldsymbol R(t),t)$.
To evaluate the TDPES $\epsilon(\boldsymbol R(t),t)$ and TDVP $\boldsymbol A_\mu(\boldsymbol R(t),t)$ along each trajectory $\boldsymbol R(t)$, one has to solve the electronic evolution equation, Eq.~(\ref{eq:eom_el}), along the same trajectory.

In order to make the CT-MQC formalism applicable for molecules, the electronic wavefunction is expressed in a basis of adiabatic electronic eigenfunctions, i.e. a Born-Huang-like expansion
\begin{equation}
\Phi_{\boldsymbol R(t)}(\boldsymbol r,t) = \sum_J C_J(\boldsymbol R(t),t)\phi_{\boldsymbol R(t)}^{J}(\boldsymbol r)
\end{equation}
and the adiabatic potential energy surfaces $\epsilon_{\text{BO}}^{J}(\boldsymbol R(t))$, i.e. the energy eigenvalues, as well as the nonadiabatic coupling vectors $\boldsymbol d_{\mu,JI}(\boldsymbol R(t))=\braket{\phi_{\boldsymbol R(t)}^J|\nabla_\mu\phi_{\boldsymbol R(t)}^I}_{\boldsymbol r}$ are introduced.

Gathering the time-dependence of the electronic wavefunction into the expansion coefficients allows us to replace the evolution equation for the electronic wavefunction with a set of ordinary differential equations for the evolution of the coefficients $\dot{C}_J(\boldsymbol R(t),t)$.
Some approximations are then introduced, along with the choice of the gauge. The main approximations are: (i) in the definition of the electron-nuclear coupling operator, Eq.~(\ref{eq:Uen}), the first term is neglected within CT-MQC\cite{eich2016adiabatic}, as well as its contribution in the expression of the TDPES, i.e. $\epsilon_{\text{geo}}$, to maintain the gauge invariance; (ii) the spatial variation of the coefficients is expressed in the form $\nabla_\mu C_J(\boldsymbol R(t),t)=\frac{i}{\hbar}\boldsymbol f_{\mu}^J(\boldsymbol R(t),t) C_J(\boldsymbol R(t),t)$, with the force for state $J$ accumulated along the trajectory $\boldsymbol f_{\mu}^J(\boldsymbol R(t),t) = - \int^t \nabla_\mu \epsilon_{\text{BO}}^J(\boldsymbol R(\tau)) d\tau$; (iii) the TDVP along a trajectory, using the Born-Huang-like expansion of the electronic wavefunction in its definition~(\ref{eq:TDVP}), is approximated as $\boldsymbol A_\mu(\boldsymbol R(t),t)=\sum_J|C_J(\boldsymbol R(t),t)|^2\boldsymbol f_\mu^J(\boldsymbol R(t),t)$, arguing that the accumulated force, which is integrated over the trajectory and thus accumulates over time, is expected to dominate over the neglected contribution that depends on spatially localized nonadiabatic coupling vectors $\boldsymbol d_{\mu,JI}(\boldsymbol R(t))$; (iv) the quantum momentum, i.e. the signature quantity of CT-MQC that naturally emerges within the exact factorization 
\begin{equation} \boldsymbol Q_\mu(\boldsymbol R(t),t)=\frac{-\hbar\nabla_\mu|\chi(\boldsymbol R(t),t)|}{|\chi(\boldsymbol R(t),t)|}
\end{equation} 
is approximated by expressing the nuclear density as a sum of frozen Gaussians centered at the positions of the trajectories, as discussed in detail elsewhere\cite{pieroni2023investigating,min2017ab}. 

With all these approximations, we arrive at the final evolution equation for the electronic coefficients along a trajectory $\boldsymbol R(t)$
\begin{align}
\dot{C}_J(\boldsymbol R(t),t)=&-\frac{i}{\hbar}\epsilon_{\text{BO}}^{J}(\boldsymbol R(t),t) C_J(\boldsymbol R(t),t) -\sum_I\sum_\mu\dot{\boldsymbol R}(t)\boldsymbol d_{\mu,JI}(\boldsymbol R(t),t) C_I(\boldsymbol R(t),t) \nonumber\\
&+\sum_\mu\frac{\boldsymbol Q_\mu(\boldsymbol R(t),t)}{\hbar M_\mu}(\boldsymbol f_\mu^J(\boldsymbol R(t),t)-\boldsymbol A_\mu(\boldsymbol R(t),t)) C_J(\boldsymbol R(t),t)
\end{align}
Recalling the characteristic equation~(\ref{eqn: TDVP as P}), we can define the classical forces for the trajectories, $\dot{\boldsymbol P}_\mu(t)=\boldsymbol F(\boldsymbol R(t),t)=\frac{d\boldsymbol A_\mu(\boldsymbol R(t),t)}{dt}$; taking the total time derivative of the approximated form of $\boldsymbol A_\mu(\boldsymbol R(t),t)$ allows us to arrive at an equivalent expression for the forces
\begin{equation}
\begin{split}
\boldsymbol F_\mu(\boldsymbol R(t),t)=&-\sum_J|C_J\boldsymbol R(t),t)|^2\nabla_\mu\epsilon_{\text{BO}}^J(\boldsymbol R(t),t) \\&-\sum_{J,I}\overline{C_J(\boldsymbol R(t),t)}C_I(\boldsymbol R(t),t)\left(\epsilon_{\text{BO}}^J(\boldsymbol R(t),t)-\epsilon_{\text{BO}}^I(\boldsymbol R(t),t)\right)\boldsymbol d_{\mu,JI}(\boldsymbol R(t),t)\\
&+\sum_J|C_J(\boldsymbol R(t),t)|^2\left(\sum_{\mu'}\frac{2\boldsymbol Q_{\mu'}(\boldsymbol R(t),t)}{\hbar M_{\mu'}}\cdot\boldsymbol f_{\mu'}^J(\boldsymbol R(t),t)\right)(\boldsymbol f_\mu^J(\boldsymbol R(t),t)-\boldsymbol f_\mu^J(\boldsymbol R(t),t))
\end{split}
\end{equation}
with $\overline{C_J}$ the complex conjugated of $C_J$. In both expressions for the evolution of the coefficients and for the force, the first two terms are identical to standard Ehrenfest dynamics; the new terms containing the quantum momentum couple the trajectories through its dependence on the nuclear density.

\section{Computational details}\label{sec: comp details}
The low-energy dynamics studied in this work are initiated in the ground-state of a  $E \otimes e$ Jahn-Teller model which was recently explored by Valahu et al.\cite{valahu2023direct}.
The electronic Hamiltonian in the diabatic basis $\lbrace|\phi^{1}\rangle,|\phi^{2}\rangle\rbrace$ reads
\begin{equation}\label{eqn: H}
\textbf{H}_{\text{el}}(X,Y)=\begin{pmatrix}\frac{\omega}{2}(X^2+Y^2) + \kappa X & \kappa Y \\ \kappa Y & \frac{\omega}{2}(X^2+Y^2) - \kappa X \end{pmatrix} 
\end{equation}
where $\kappa=2\pi$kHz, $\omega=4\pi/3$kHz. $\boldsymbol{R}=(X,Y)$ are the nuclear coordinates. The nuclear kinetic energy operator is $\hat{T}_{\text{n}}=\omega(\hat{P}_X^2+\hat{P}_Y^2)/2$.
At time $t=0$ the function $\chi(\boldsymbol{R},0)$ is a real Gaussian centered at $\boldsymbol R_0=(-2.5,0)$ with variance 1 to ensure $|\chi(\boldsymbol{R},0)|^2\approx|\chi_{S_0}(\boldsymbol{R},0)|^2$. 
 Exact propagation is performed in the diabatic basis using the split-operator technique~\cite{feit1982solution} for 1.5 ms with a time step of 10 ns. A grid consisting of 501 grid points in $X$ and $Y$ is used in the range $X,Y \in [-8,8]$.
The TDVP and TDPES are reconstructed from the diabatic quantities as detailed in Ref.~[\citenum{ibele2022photochemical}].

Based on the TDPES and TDVP, 500 quantum and classical trajectories are propagated with initial positions  sampled from the initial probability density of the nuclear wavefunction. For the classical trajectories, positions and momenta are treated as independent variables and the momenta are sampled from the momentum probability distribution from the Wigner transform of the initial nuclear wavefunction. The momenta of the quantum trajectories at a certain time $t$ and position $\boldsymbol R$ are given by the TDVP $\boldsymbol A(\boldsymbol R,t)$. Consequently, the initial momenta corresponding to initial positions $\boldsymbol R_0$ are obtained as $\boldsymbol A(\boldsymbol R_0,0)$. A time step of 1 $\mu$s is used and the stored TDPES and TDVP are used at each time step.

For CT-MQC, the 500 Wigner sampled trajectories (positions and momenta) are propagated using the G-CTMQC software.\cite{gctmqc} A nuclear and electronic integration time step of 10 ns is used. 

\section{Results}\label{sec: results}
We discuss in this section the results of the quantum wavepacket dynamics from the model Hamiltonian of Eq.~(\ref{eqn: H}) adopting the adiabatic representation. In Section~\ref{sec:nonadia}, instead, we will discuss the dynamics in the eye of the exact factorization, but we will relate our analysis to the dynamics described here.

\begin{figure}
\centering
\includegraphics[width=\linewidth]{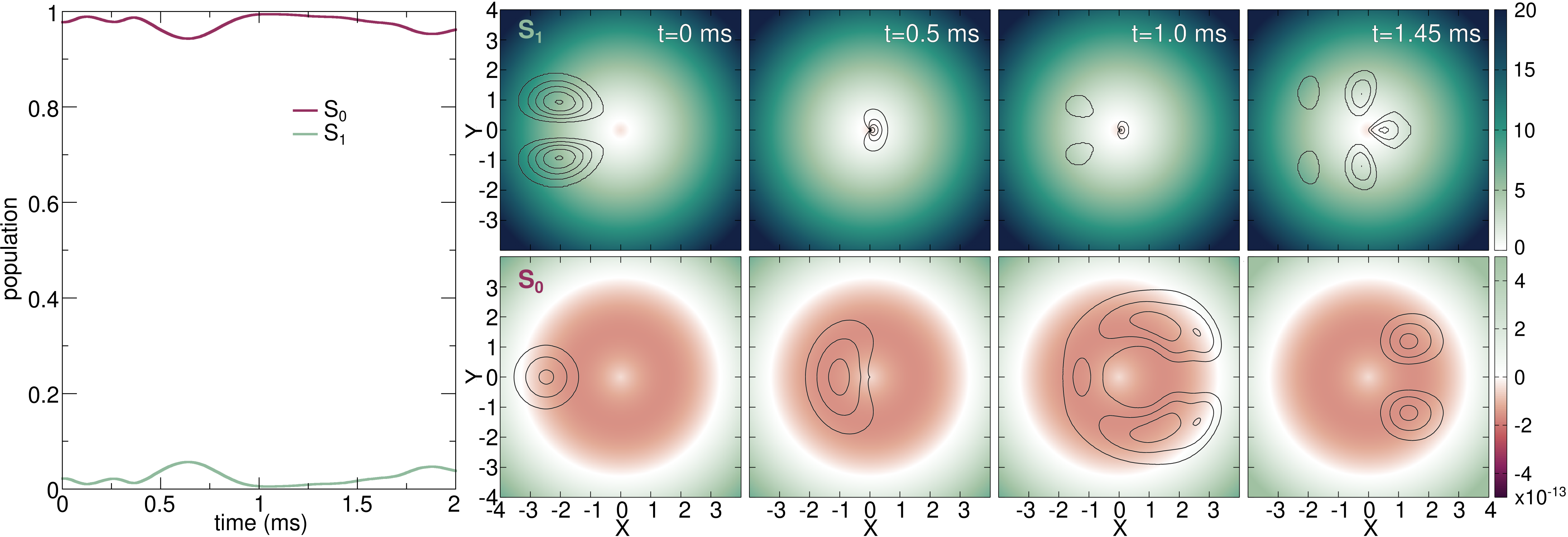}
\caption{Dynamics in the adiabatic representation. Left panel: populations of the adiabatic states S$_0$ (brown) and S$_1$ (green) as functions of time. Right panels, top: adiabatic excited-state potential energy surface represented as colormaps with the black contour lines representing the excited-state contribution to the total density at the four representative snapshots indicated in the panels. Right panels, bottom: adiabatic ground-state potential energy surface represented as colormaps with the black contour lines representing the ground-state contribution to the total density at the same four representative snapshots.} \label{fig:adia_pict}
\end{figure}

As discussed in Section~\ref{sec: comp details}, the simulated low-energy dynamics starts with the nuclear density centered around $\boldsymbol R_0=(-2.5,0)$, thus $X<0,Y=0$, and evolves towards the conical intersection at $\boldsymbol R_{\text{CI}}=(0,0)$, reaching the region $X>0,Y=0$. Such low-energy dynamics remain mostly in the adiabatic ground state S$_0$, as confirmed by the populations of the two electronic states plotted as functions of time in the left panel of Fig.~\ref{fig:adia_pict}. However, during the whole simulation, we observe that up to 5\% of population is transferred to the excited state S$_1$, confirming that nonadiabatic effects are weak, but not negligible. This is due to the fact that the at conical intersection the coupling between the electronic states is singular, and manifests itself as a ``reverted funnelling'' process from S$_0$ to S$_1$.

The adiabatic potential energy surfaces are shown in Fig.~\ref{fig:adia_pict} in the right panels, with S$_1$ in reported at the top and S$_0$ at the bottom as colormaps. The ground state presents a Mexican-hat-like shape with a cusp at the center at the position of the conical intersection, while the excited state is basically a well that terminates at the bottom with a cone shape whose vertex is degenerate with the cusp of the ground state: there, the typical double-cone shape at the conical intersection appears clearly. The four panels on the right of Fig.~\ref{fig:adia_pict} show the same (static) potentials but at the different time snapshots we observe the ground-state and the excited-state contributions to the nuclear density that evolve from $X<0$ towards $X>0$ remaining symmetric around $Y=0$.

The adiabatic nuclear densities, after having been initialized around $\boldsymbol R_0$ with the majority of the population in the ground state, move towards the conical intersection. At $t=0.5$~ms, the ground-state density transfers some population to the excited state, and density appears in the direct vicinity of the conical intersection. Note that ground-state density is completely absent at $\boldsymbol R_{\text{CI}}$ because it has been funnelled to the excited state. In moving towards $X>0,Y=0$, at $t=1.0$~ms, the ground-state density branches in the positive and negative directions along $Y$, remaining (numerically) zero for $X>0,Y=0$. This behavior is the manifestation of destructive interference during the evolution of the S$_0$ wavepacket encircling the conical intersection, but, as observed in our previous study~\cite{ibele2023nature}, it does not affect the full nuclear density, only its ground-state component. At the conical intersection, non-zero density appears only in the excited state. Finally, at $t=1.45$~ms, while the dynamics progresses towards $X>0$, we note that the region $X>0,Y=0$ is only populated by nuclear density in the excited state.

\subsection{Nuclear dynamics through quantum and classical trajectories}\label{sec:nonadia}

\begin{figure}
\centering
\includegraphics[width=\textwidth]{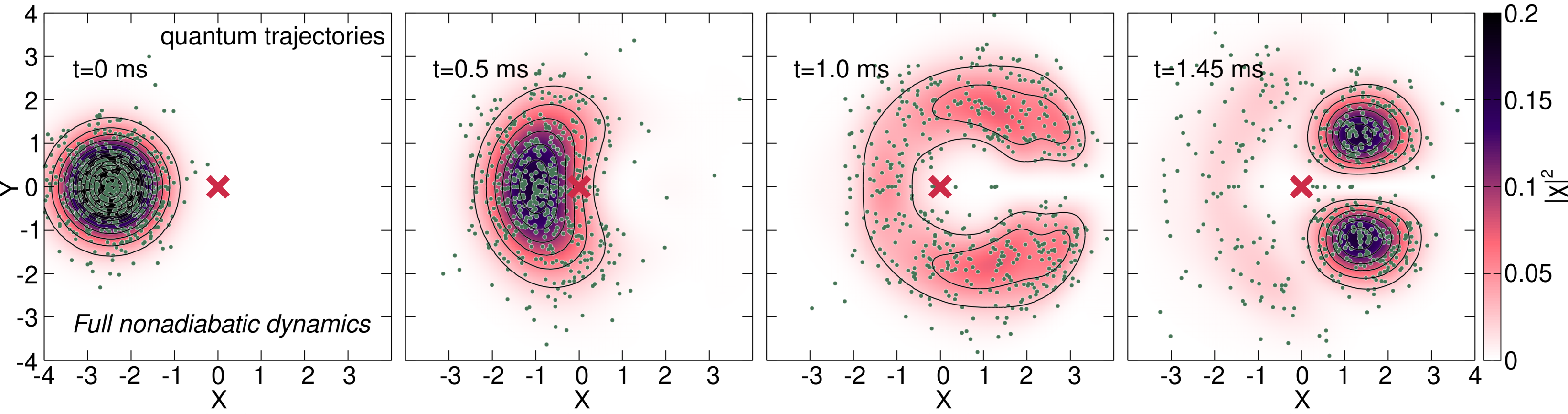}
\caption{The nuclear density at four snapshots along the dynamics (colour map and contour lines). The green dots indicate the positions of the quantum trajectories at the same time step. $\boldsymbol R_{\text{CI}}$ is indicated by the red cross.} \label{fig:ndens}
\end{figure}
In order to investigate the low-energy dynamics around the region where the conical intersection is located in the adiabatic representation, we first look at the evolution of the full nuclear density.
In Fig.~\ref{fig:ndens}, we show the nuclear density at four snapshots throughout the dynamics. At $t=0$ ms, the dynamics is initialized centered around $\boldsymbol R_0=(-2.5,0)$. At $t=0.5$ ms, the nuclear density reaches the point of the conical intersection at $\boldsymbol R_{\text{CI}}=(0,0)$, indicated by a red cross in all plots. At $t=1.0$ ms, the wavepacket evolves towards positive $X$ values, splitting in $Y$ direction symmetrically around $Y=0$. Finally, at $t=1.45$ ms, two parts of the wavepacket have formed that evolve at positive $X$ values, while a diffuse part of the wavepacket remains spread over $X<0$. 
\begin{figure}
\centering
\includegraphics[width=\textwidth]{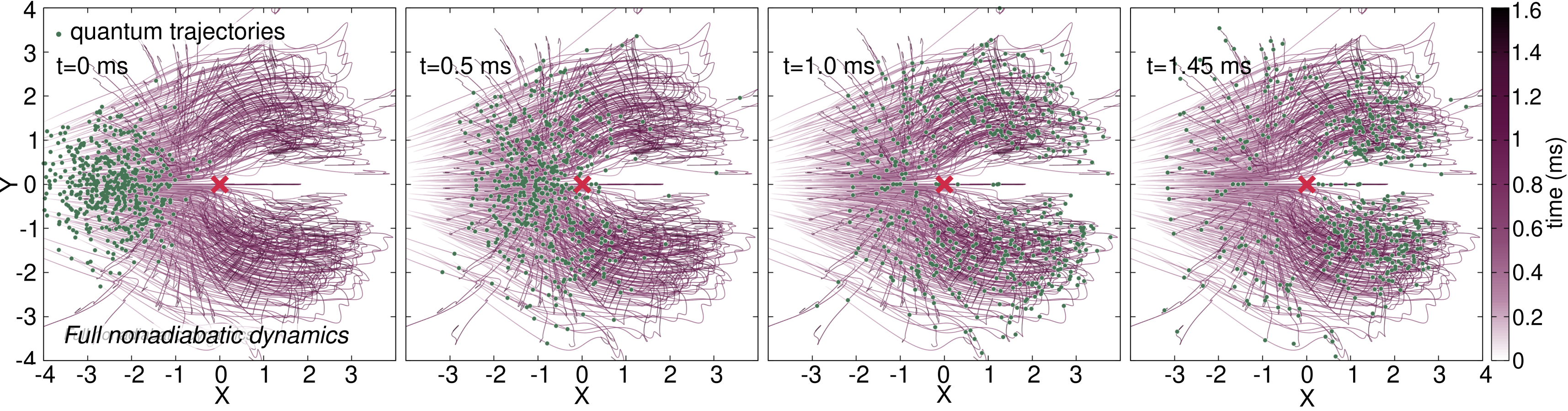}
\caption{Snapshots of the positions of the quantum trajectories indicated by green dots propagated along the full nonadiabatic dynamics. The overall time traces of the quantum trajectories are shown in each plot with the time indicated by the colour bar. $\boldsymbol R_{\text{CI}}$ is indicated by the red cross.} \label{fig:quant_full}
\end{figure}

Subsequently, we compare the wavepacket evolution with the dynamics of the quantum trajectories. Quantum trajectories are by construction expected to reproduce closely the quantum dynamics, and in the limit of an infinite amount of trajectories, they would converge to the same dynamics.
The quantum trajectories at all snapshots in Fig.~\ref{fig:ndens} follow closely the evolution of the nuclear density. The most interesting observation is that along and in close vicinity of $Y=0$, there are indeed several quantum trajectories evolving past $X=0$. This clearly indicates the presence of nuclear density in this region of configuration space, which supports our findings of Ref.~[\citenum{ibele2023nature}] against the formation of a zero-density line, which would appear as a consequence of the effect of topological-phase interferences in relation to the conical intersection.

The behaviour of the quantum trajectories can be clearly understood by plotting the traces of the trajectory and specific positions at the same four snapshots discussed above (Fig.~\ref{fig:quant_full}).
 The colour of the trace indicates the time. The trajectories, randomly sampled from the initial density, are initially spread within $[-4:0]$ in the $X$ direction and $[-2:2]$ in the $Y$ direction. Once they evolve towards $\boldsymbol R_{\text{CI}}$ ($t=0.5$ ms), there is a clear trend of the majority of trajectories avoiding $X>0, Y=0$ visible through the deviation of the traces towards positive and negative $Y$ values. However, a small number evolves unfaced through the region of $\boldsymbol R_{\text{CI}}$, very close to and at $Y=0$, without any deviation in the $Y$ direction; these trajectories keep evolving solely in the $X$-direction throughout the dynamics, and the time traces validates this observation. Thus, in general, there is a tendency of avoiding the configuration space around $Y=0$ for $X>0$ resulting in a low density of trajectories in the vicinity of this region. However, this region is not completely unaccessible, as shown by the non-negligible number of the trajectories evolving there.
These quantum trajectories strongly underline the fact that in the region of configuration space around $Y=0$ for $X>0$ the nuclear density is non-zero. This is of particular interest, since, under the effect of a quantized topological phase, one would expect destructive interference in the quantum wavepacket, leading to a zero-density line in the nuclear density and, subsequently, in the trajectory distribution. However, this is clearly not the case.

While the quantum trajectories reproduce all the significant features of the nuclear-density dynamics, we can evaluate the effect that the classical approximation, i.e. neglecting the quantum potential, has on these complex dynamics. 
This topic has been previously investigated for adiabatic dynamics as well as excited state dynamics.\cite{Scribano_JCTC2022, ibele2022photochemical} 
However, to the best of our knowledge, few studies have looked into this topic for low-energy dynamics where nonadiabatic effects still come into play in the vicinity of the region where the conical intersection appears in the adiabatic representation. 
Note that a major part of simulations of molecular adiabatic and nonadiabatic dynamics rely on a classical trajectory framework. Therefore, it is highly relevant to understand the direct limitations of this approximation and estimate if it is appropriate to describe low-energy dynamics at conical intersections.

\begin{figure}
\centering
\includegraphics[width=\textwidth]{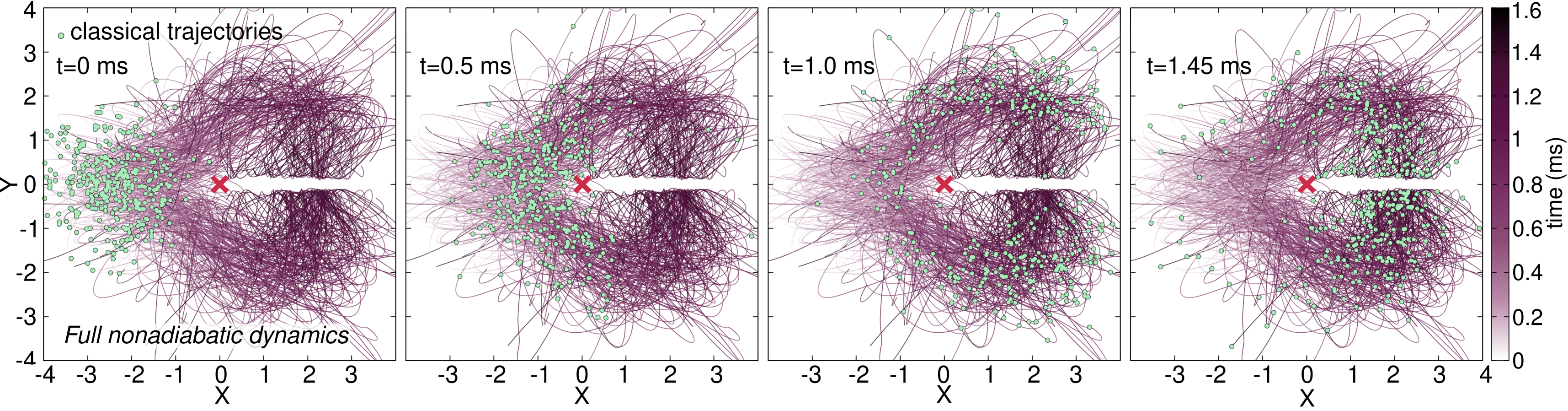}
\caption{Snapshots of the positions of classical trajectories indicated by green dots propagated along the full nonadiabatic dynamics. The overall time traces of the classical trajectories are shown in each plot with the time indicated by the colour bar. $\boldsymbol R_{\text{CI}}$ is indicated by the red cross.} \label{fig:class_full}
\end{figure}

Figure~\ref{fig:class_full} depicts the time traces and position of classical trajectories propagated based on the exact TDVP and TDPES but without the quantum potential. The positions of the trajectories give an overall similar picture as the quantum trajectories and thus, reasonably well reproduce the quantum dynamics. Nonetheless, when looking at the time traces, qualitative differences arise: The classical trajectories completely avoid the region of $Y \in [-0.1,0.1]$ for $X>0$. 
The majority of trajectories approaches this region at different times later during the dynamics, but these trajectories are seemingly repelled. Especially at $t=1.45$ ms, a large number of trajectories have approached $Y=0$ for $X>0$, but their momentum is being reverted in $Y$ direction. 

\begin{figure}
\centering
\includegraphics[width=\textwidth]{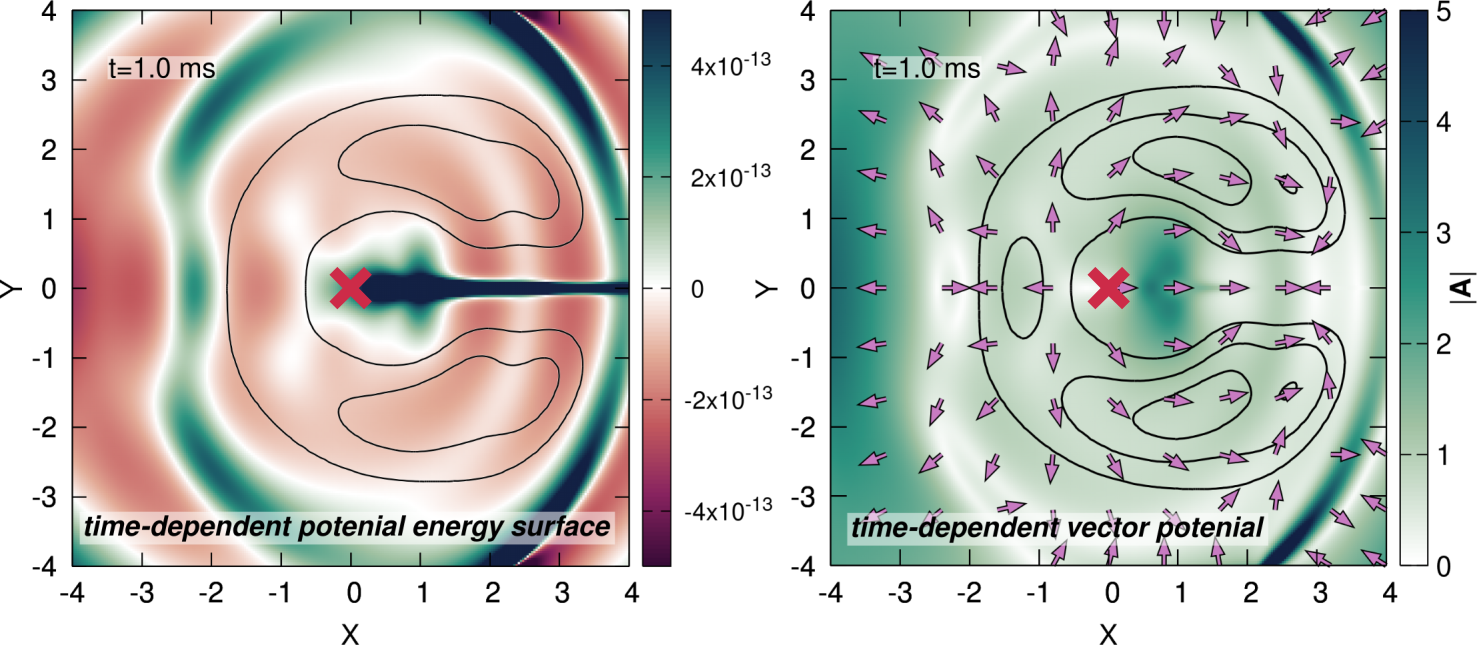}
\caption{Left panel: The time-dependent potential energy surface (TDPES) at $t=1.0$ ms is indicated by the colour map. Right panel: The absolute value of the time-dependent vector potential (TDVP) at $t=1.0$ ms is indicated by the colour map and its directionality is given by a unit vector.
In both panels, the nuclear density at the same time step is indicated by the contour lines and $\boldsymbol R_{\text{CI}}$ is indicated by the red cross.} \label{fig:tdpes_tdvp}
\end{figure}

In order to explain this behaviour and its origin, we turn to the time-dependent potentials of the exact factorization, the TDPES and TDVP.
It is important to recall the difference between the quantum and classical trajectory frameworks. As explained in Section~\ref{sec:traj}, the TDVP is the nuclear momentum field and is used in our quantum trajectory calculations to drive the evolution of the trajectories. For the classical trajectories, the TDPES and its gradients are used to reconstruct the nuclear momenta and, thus, drive the trajectories.  Figure~\ref{fig:tdpes_tdvp} shows the TDPES and TDVP at a representative time step  $t=1.0$ ms. The TDPES (left plot in Fig.~\ref{fig:tdpes_tdvp}) forms a strong barrier along $X>0,Y=0$ that has some finite width in the $Y$ direction for all $X>0$. This barrier persists in time and it is responsible for the classical trajectories not being able to access the region $X>0,Y=0$. 
In comparison, the TDVP does not show a particular feature along $Y=0$. Due to the symmetry of the potential and the initial wavefunction, the $Y$ component of the TDVP is zero along $Y=0$ which explains the motion of the quantum trajectories in the vicinity of $Y=0$ that move strictly along the $X$ direction all along the dynamics. 

\begin{figure}
\centering
\includegraphics[width=\textwidth]{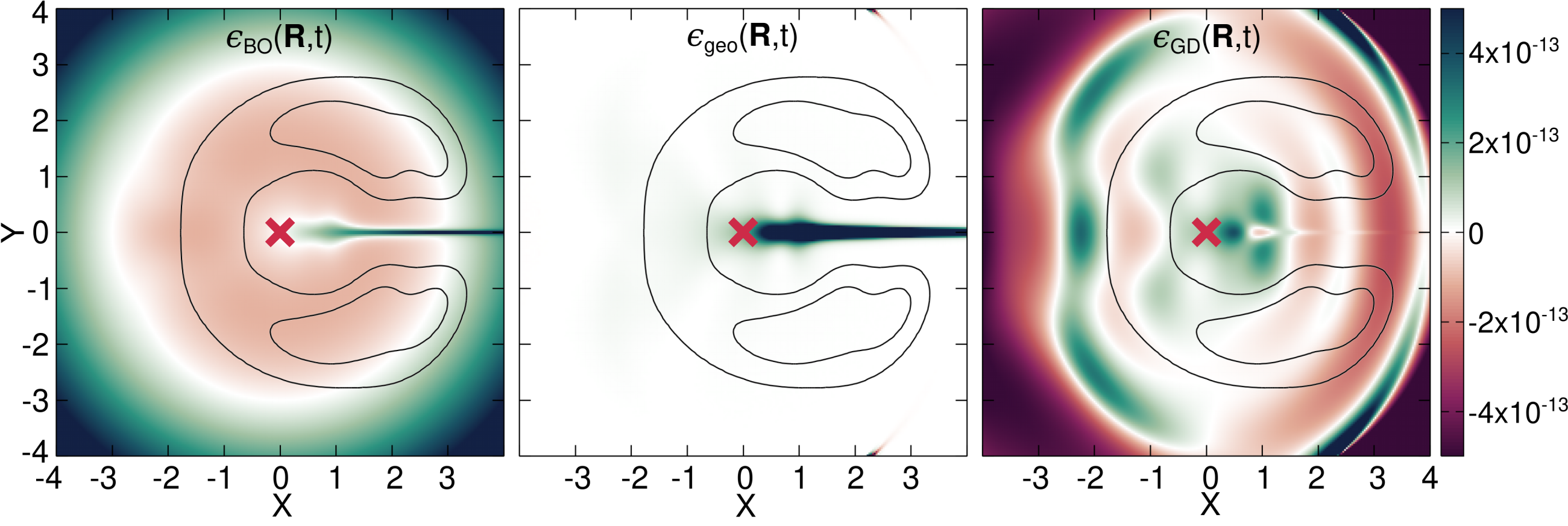}
\caption{The three terms contributing to the time-dependent potential energy surface (TDPES) at time $t=1.0$~ms. In all panels, the nuclear density at the same time step is indicated by the contour lines and $\boldsymbol R_{\text{CI}}$ is indicated by the red cross. Left panel: Born-Oppenheimer term $\epsilon_{\text{BO}}$, central panel: geometric term $\epsilon_{\text{geo}}$, right panel: gauge-dependent term $\epsilon_{\text{GD}}$.} \label{fig:tdpes}
\end{figure}

To understand in more detail the barrier formed in the TDPES, we plot in Fig.~\ref{fig:tdpes} the three contributions to the TDPES at the representative time step $t=1.0$ ms as given in Eqs.~(\ref{eq:tdpes_bo}), (\ref{eq:tdpes_geo}), and (\ref{eq:tdpes_gd}): The Born-Oppenheimer term, $\epsilon_{\text{BO}}$, in the left panel, the geometric term, $\epsilon_{\text{geo}}$, in the central panel, and the gauge-dependent term $\epsilon_{\text{GD}}$, in the right panel.
We observe a different manifestation of this barrier in all three terms. The term $\epsilon_{\text{BO}}$ encodes, as expected from its definition, the overall shape of the TDPES that closely resembles the potential energy surface of the adiabatic ground state (see Fig.~\ref{fig:adia_pict} for comparison), where the main part of the dynamics takes place. If one were to decompose into contributions from the adiabatic ground state and from the adiabatic excited state the shape of $\epsilon_{\text{BO}}$ at $t=1.0$~ms, one would associate only the region around $\boldsymbol R_{\text{CI}}$ and along $X>0, Y=0$ to the excited state, because in this region the ground-state wavepacket disappears, as it is transferred to the excited state in a ``reverted funnelling'' process.
This yields a small barrier that is closely localized around $Y=0$ for $X>0$. 
The geometric contribution, $\epsilon_{\text{geo}}$, has a very peculiar behavior, as it is almost zero in the majority of the configuration space. However, in the region close to $\boldsymbol R_{\text{CI}}$, where one expects mixing of the two adiabatic states involved, $\epsilon_{\text{geo}}$ contributes significantly to the TDPES with a barrier. Most importantly, this is the main contribution to the observed wide barrier along $Y=0$ for $X>0$ in the full TDPES of Fig.~\ref{fig:tdpes_tdvp} (left panel). Thus, the depletion in nuclear density at long times at $X>0,Y=0$ can be traced back to be largely caused by this contribution. 
Finally, the gauge-dependent contribution, $\epsilon_{\text{GD}}$, oscillates over the configuration space and seems to form a small barrier which confines the density in $X$ and $Y$ direction and prevents it from delocalizing further in configuration space. Additionally, it also forms a barrier around $\boldsymbol R_{\text{CI}}$.

\begin{figure}
\centering
\includegraphics[width=\textwidth]{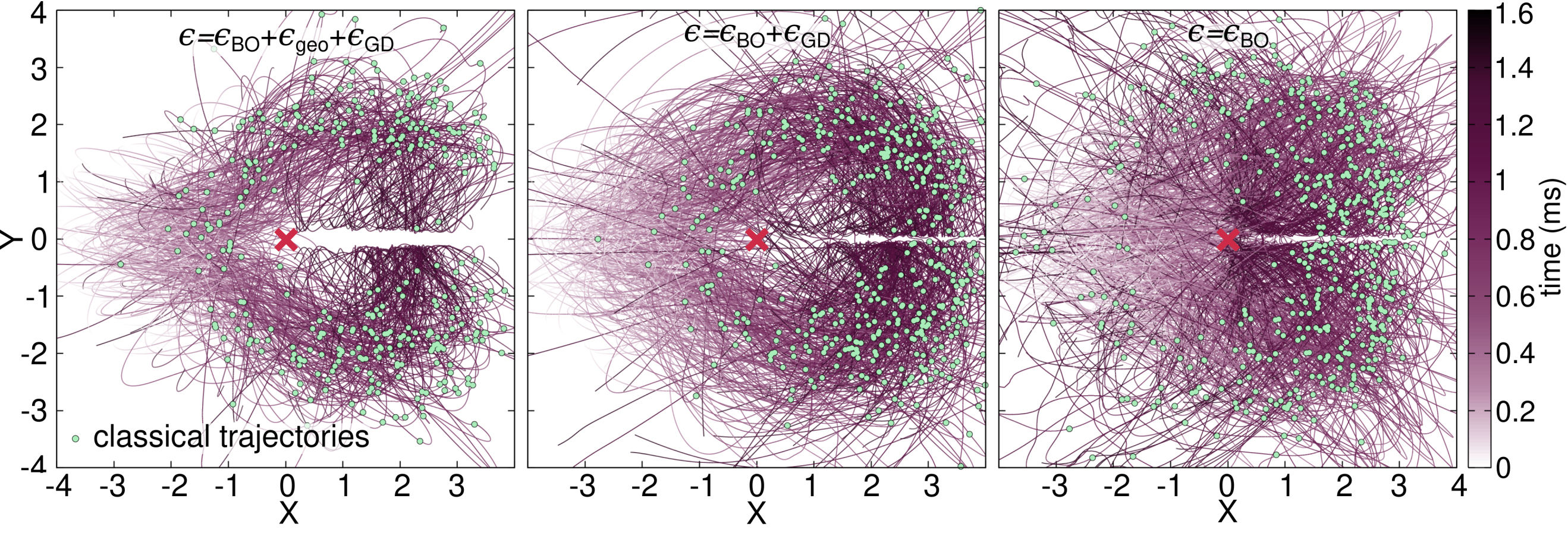}
\caption{Snapshots of the positions of classical trajectories at time $t=1.0$ ms indicated by green dots using different terms of the TDPES. Left panel: Full TDPES, central panel: partial TDPES constructed with only Born-Oppenheimer and gauge-dependent terms, right panel: only Born-Oppenheimer term. 
The overall time traces of the classical trajectories are shown in each plot with the time indicated by the colour bar. $\boldsymbol R_{\text{CI}}$ is indicated by the red cross.} \label{fig:class_rt}
\end{figure}

Following from these observations of the TDPES, the question arises as to how these contributions affect the classical trajectories over the full course of the dynamics. 
In the approximate trajectory schemes derived from the exact factorization,\cite{min2017ab,pieroni2023investigating,ha2018surface} some of these or similar contributions to the dynamics are neglected or approximated. Therefore, by providing a deeper understanding of the effect that each term has on a classical trajectory-based dynamics, we can offer insight into which approximations deserve particular care.

All these terms are time dependent, so looking at just one specific snapshot, although informative, is insufficient to estimate entirely its contribution and importance for the evolution of classical trajectories.
Therefore, we repeated the same dynamics for classical trajectories progressively removing terms of the TDPES. 
In Fig.~\ref{fig:class_rt}, we show the snapshot of the trajectories at time $t=1.0$ ms as well as the time traces of each trajectory. The left panel reports again, for comparison, the dynamics on the full TDPES that clearly shows the trajectories avoiding a large region around $X>0,Y=0$. In the central panel of Fig.~\ref{fig:class_rt}, we show the dynamics without the geometric contribution $\epsilon_{\text{geo}}$. From the positions of the trajectories, it can already be seen that the dynamics are sped up with respect to the full TDPES. The most significant feature becomes clear when looking at the time traces: the trajectories still mostly avoid the region  $X>0, Y=0$ but they approach significantly closer the line $Y=0$ and several individual trajectories even cross through $Y=0$.
This brings the dynamics qualitatively closer to what is observed in the quantum trajectories, although they now overestimate how close the trajectories can approach $Y=0$. We can therefore argue that the absence of classical trajectories at $X>0$, $Y=0$ is majorly caused by the $\epsilon_{\text{geo}}$ contribution. It appears that the quantum potential counteracts the effect of $\epsilon_{\text{geo}}$, thus, without the quantum potential, one gets spurious zero-density region at $X>0$, $Y=0$. However, $\epsilon_{\text{geo}}$ cannot simply be neglected as it is crucial to describe the nuclear density/trajectories surrounding $X>0$, $Y=0$.
When also the gauge-dependent term, $\epsilon_{\text{GD}}$, is removed, i.e. using only $\epsilon_{\text{BO}}$, (Fig.~\ref{fig:class_rt}, right panel) the trajectories approach closely $\boldsymbol R_{\text{CI}}$ and there is almost no repulsion visible in the region surrounding $\boldsymbol R_{\text{CI}}$. The trajectories still avoid a very small portion of space around $X>0,Y=0$ and the dynamics are even further sped up with the trajectories already moving back towards $X<0$ at 1.0 ms. 

Summarizing, we see substantial differences between the quantum and classical trajectories in particular when focusing on the region of low nuclear density along $X>0,Y=0$. Both sets of trajectories are able to capture the small nuclear density in this region, but the classical trajectories overestimate this effect.

The TDPES shows a fairly wide, finite barrier along $Y=0$ for $X>0$, which is predominantly caused by the $\epsilon_{\text{geo}}$ contribution. As this barrier is finite in height, it can be expected that in the limit of an infinite number of trajectories, this region of configuration space will be populated. 

Recalling the above presentation of the wavepacket dynamics decomposed into ground-state and excited-state contributions, we observe that the density transferred through the conical intersection to the first excited state is the density that evolves from the negative to the positive $X$ ``undisturbed'', i.e. with only a negligible momentum contribution in $Y$ direction, in close vicinity of $Y=0$. As the quantum trajectories are propagated in the exact-factorization framework, i.e. only based on the TDPES and the TDVP, such a direct relation to the adiabatic contributions is not possible. However, we can assume that the trajectories accessing the configuration space around $X>0,Y=0$ can be ``attributed'' to the adiabatic excited state: as the nuclear density  evolves towards positive $X$ with a contribution of the momentum field only along $X$, as seen from the wavepacket propagation, so the quantum trajectories evolving in this part of configuration space have zero momentum along $Y$. During the wavepacket dynamics, up to 5\% of population is accessing the excited state, which is qualitatively well reproduced by the fraction of quantum trajectories evolving in $X>0,Y=0$, while largely underestimated by the classical trajectories.  
Thus, it can be argued that the quantum potential is essential to ensure a proper description of these dynamics in the trajectory framework. 

\subsection{CT-MQC for a low-energy process around $\boldsymbol R_{\text{CI}}$}
As we have seen the classical trajectories being strongly influenced by the terms of the TDPES that are included in the dynamics, we explore in this section the fully approximate CT-MQC approach, where the TDPES is constructed on-the-fly for each trajectory from the adiabatic quantities. This approach is mainly targeted for the use on molecular systems. Its performance has been estimated for molecular systems\cite{min2017ab} and for a large variety of low dimensional model systems recreating the deactivation through a conical intersection~\cite{Agostini_JCP2022, Agostini_JCP2021_1, Agostini_JCTC2020_2}. However, so far, its capabilities of describing low-energy dynamics which exhibit some nonadiabatic behaviour has not been tested. This is not a pure theoretical curiousity but also a relevant problem for real molecular systems.\cite{kendrick1996geometric,desouter1983transition,carpenter2022conical}

\begin{figure}
\centering
\includegraphics[width=\textwidth]{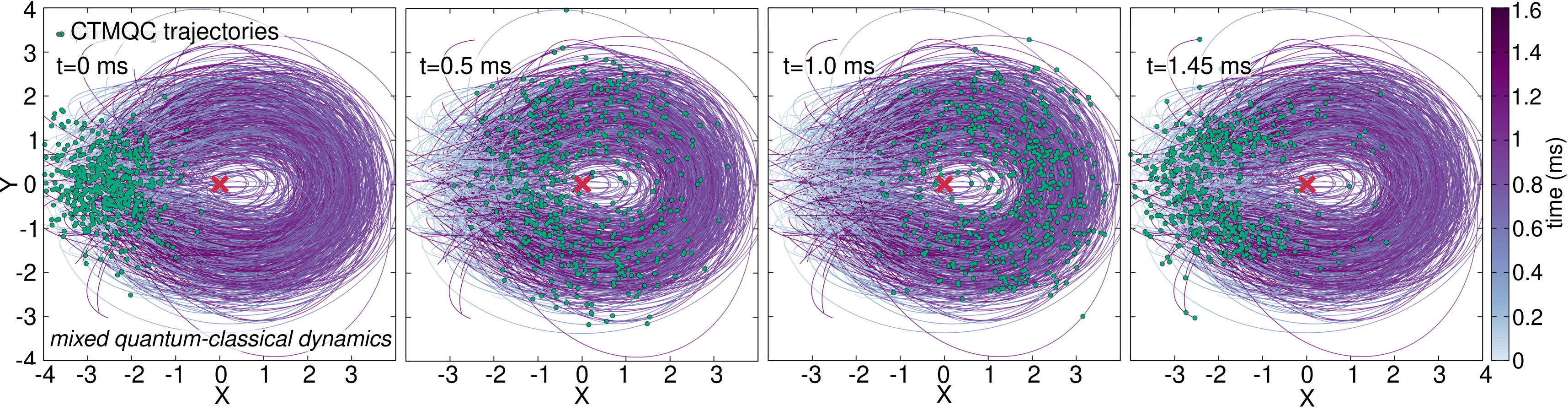}
\caption{Snapshots of the positions of classical CT-MQC trajectories indicated by green dots. The overall time traces of the CT-MQC trajectories are shown in each plot with the time indicated by the colour bar. $\boldsymbol R_{\text{CI}}$ is indicated by the red cross.} \label{fig:ctmqc}
\end{figure}

In Fig.~\ref{fig:ctmqc} we depict, as done previously, the positions of the trajectories at four snapshots of the dynamics overlaid with the time traces of all the trajectories. 
The trajectories are not showing any repulsion along $X>0,Y=0$ and they evolve smoothly surrounding $\boldsymbol R_{\text{CI}}$ without avoiding the region $X>0, Y=0$. 
This behavior can be attributed to the fact that the simulation is performed in the adiabatic basis (the algorithm, and thus the underlying approximations, is derived in the adiabatic representation). First, we observe from the calculation of the populations of the adiabatic states (not shown) that the trajectories do not feel the nonadiabatic effect in the vicinity of the CI and all the coefficients of the adiabatic excited state remain zero throughout the dynamics. Thus, since in CT-MQC one reconstructs the TDPES and the TDVP through the adiabatic quantities, the equations of motion derived in Section~\ref{sec:ctmqc} show that in the case of zero population transfer, the force acting on the trajectories reverts back to the negative gradient of the adiabatic energy. This implies that the trajectories are classically evolving only according to the adiabatic ground state. Due to the shape of the adiabatic ground state PES, the trajectories evolve surrounding $\boldsymbol R_{\text{CI}}$ but there is no hint towards an energy barrier for $X>0,Y=0$ as in the TDPES. This analysis explains why CT-MQC classical trajectories behave qualitatively differently from the classical trajectories propagated using the full TDPES. On the other hand, also those classical trajectories did not feel the nonadiabatic effect encoded in the TDPES through $\epsilon_{\text{geo}}$: none of the 500 trajectories of Fig.~\ref{fig:class_full} felt such a nonadiabatic effect, if we assume that it is only localized in the region $X>0,Y=0$ where $\epsilon_{\text{geo}}$ is strongly non-zero.


In addition, the trajectories are evolving significantly faster than the quantum and classical trajectories propagated on the TDPES. Specifically, after 1.45 ms the majority of the trajectories is back in the area of initialization. In the classical trajectories (Section~\ref{sec:nonadia}), neglecting the gauge-dependent and geometric terms of the TDPES caused a significantly faster evolution of the trajectories. The same is true for the CT-MQC trajectories, where the these terms are in part neglected (geometric term) and in part approximated (gauge-dependent).

In order to differentiate the effects of the classical approximation and of the reconstruction of the TDPES and TDVP using adiabatic quantities, we perform numerically exact quantum dynamics simulations on the adiabatic ground state potential uncoupled from the excited state. These dynamics can be subsequently analyzed further using quantum and classical trajectories, in analogy to Section \ref{sec:nonadia}.

\begin{figure}
\centering
\includegraphics[width=\textwidth]{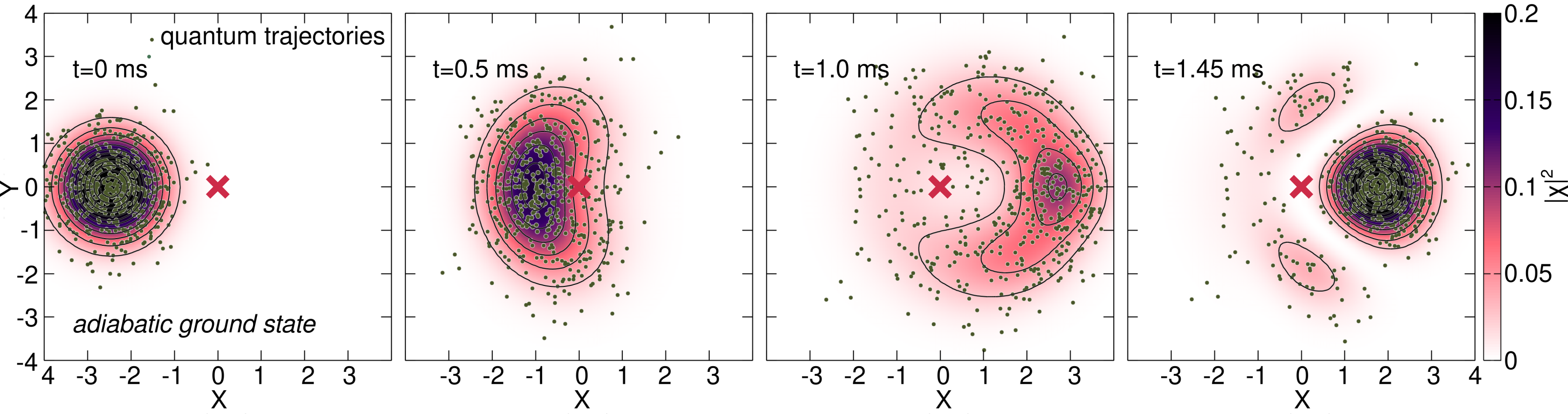}
\caption{The nuclear density at four snapshots along the adiabatic ground-state dynamics (colour map and contour lines). The green dots indicate the positions of the quantum trajectories at the same time step. $\boldsymbol R_{\text{CI}}$ is indicated by the red cross.} \label{fig:adia_q}
\end{figure}

In Fig.~\ref{fig:adia_q}, we show this ground-state evolution of the nuclear density. The initial evolution up to 0.5 ms looks very similar to the fully nonadiabatic case, with the wavepacket approaching $\boldsymbol R_{\text{CI}}$ and starting to get diverted towards positive and negative $Y$ values. After 1 ms, the wavepacket has split into two equal parts in positive and negative $Y$, but these parts recombine in the positive $X$ region showing constructive interference along $Y=0$. The area around $\boldsymbol R_{\text{CI}}$ is less densely populated than the $X>0,Y=0$ region, however, there is still a significant population there. Finally, after 1.45 ms, the wavepacket starts moving back towards $X<0$, with the larger amount of density centered along $Y=0$ for $X>0$ and with a smaller contribution, that split towards positive and negative $Y$ values, that moves backs towards $X<0,Y=0$.

The quantum trajectories of Fig.~\ref{fig:adia_q} follow closely this ground-state adiabatic evolution of the nuclear density. Most importantly, the trajectories evolve in the vicinity of $\boldsymbol R_{\text{CI}}$. At $t=1.0$ ms, for instance, when the nuclear density is reduced in that region of configuration space, there is still a non-negligible number of trajectories present.
\begin{figure}
\centering
\includegraphics[width=\textwidth]{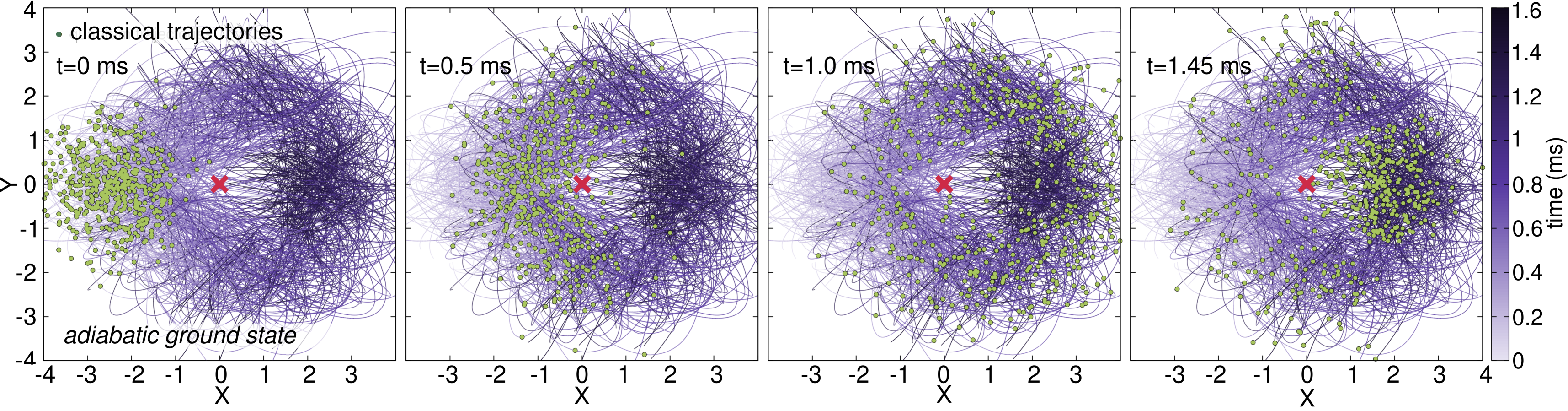}
\caption{Snapshots of the positions of classical trajectories indicated by green dots propagated along adiabatic ground-state dynamics. The overall time traces of the classical trajectories are shown in each plot with the time indicated by the colour bar. $\boldsymbol R_{\text{CI}}$ is indicated by the red cross.} \label{fig:adia_c}
\end{figure}
The classical trajectories, as depicted in Fig.~\ref{fig:adia_c}, are distributed similarly to the quantum nuclear density. Following the time traces, it can be seen that the area around $\boldsymbol R_{\text{CI}}$ is not entirely avoided, however, there are not many trajectories passing through, which is in line with the reduced nuclear density observed in the quantum propagation. 
In general, we note that in the adiabatic propagation, the classical approximation does not significantly affect the dynamics. Both, qualitatively and quantitatively, the propagation of the classical trajectories agrees well with the quantum trajectories as well as the wavepacket propagation. 

Comparing the classical, adiabatic trajectories with the CT-MQC trajectories, we observe that the overall dynamics shows several similarities. In particular, the trajectories evolve in a circular-like shape around $\boldsymbol R_{\text{CI}}$, while slightly avoiding the region close to $\boldsymbol R_{\text{CI}}$ for $X>0$. There are two main differences that the trajectories exhibit: 1) as demonstrated by the different time traces reported in Figs.~\ref{fig:ctmqc} and~\ref{fig:adia_c}, the classical, adiabatic trajectories do not form an elliptic smooth shape around $\boldsymbol R_{\text{CI}}$, as done, instead, by the CT-MQC trajectories; 2) the dynamics are significantly faster with CT-MQC, as CT-MQC trajectories return to the region of configuration space where they were initialized already after 1.45 ms, while the classical, adiabatic trajectories are still predominantly located at $X>0$. 
The faster time scale, however, can be explained by the approximations underlying the CT-MQC algorithm, that neglect the geometric term of the TDPES: the results presented in Fig.~\ref{fig:class_rt} shoed that removing this contribution from the TDPES leads to a significant speed up in the trajectories dynamics.

Overall, we conclude that the classical approximation has to be used delicately to describe this low-energy dynamics, and, due to the missing description of nonadiabatic effects, the approximated CT-MQC dynamics is reverted to a purely adiabatic process unaffected by the presence of the conical intersection.

\section{Conclusion}\label{sec: concl}
In this work, we investigated different trajectory frameworks to describe low-energy dynamics in the vicinity of a conical intersection (that appears in the adiabatic representation of the dynamics). The particularly challenging nature of this process lies within the fact that nonadiabatic effects are weak and very localized in configuration space, but significantly alter the observed dynamics. The comparison between quantum wavepacket dynamics and quantum trajectories propagated on the TDPES (and with the quantum potential) confirms previous observations on the absence of topological phase effects, that would manifest themselves with a zero-density line in $X>0,Y=0$ in the present study, due to destructive interferences when two branches of the wavepackets/trajectories move around the position where the conical intersection is located in the adiabatic representation. Nonetheless, quantum trajectories are able to accurately capture the depletion of density in the same region where the zero-density line would be expected.

We tested the performance of the classical approximation by employing two frameworks of classical trajectories, and both have difficulties to accurately reproduce nonadiabatic effects. The classical trajectories propagated with the exact TDPES (without the quantum potential) actually completely avoid the region of configuration space $X>0,Y=0$, thus manifesting a behavior reminiscent of the destructive interference related to the topological phase. This occurs because the nonadiabatic effects that counteract the formation of the zero-density line are underestimated. Thus, in this dynamics, it is the quantum potential that ensures a proper description of the nonadiabatic nature of the dynamics. Dissecting the contributions of the different terms of the TDPES has shown that the $\epsilon_{\text{geo}}$ term as well contributes to a proper account of nonadiabatic effects. The classical trajectories propagated according to the CT-MQC algorithm and, thus, according to an approximated TDPES, do not exhibit any nonadiabatic effects. Therefore, the dynamics reduces to a fully classical propagation along the adiabatic ground state. These dynamics entirely fail to reproduce the influence of the conical intersection: there is no depletion of density in the region $X>0,Y=0$ once the trajectories pass the conical intersection. 

Finally, we propagated quantum dynamics adiabatically on the ground state and obtained the numerically exact TDPES. The trajectories evolved according to such a TDPES show, indeed, a similar behavior as the CT-MQC trajectories. However, the lack of the geometric contribution to the TDPES and the approximate nature of the gauge-dependent term in CT-MQC, together with the complete absence of nonadiabaticiy, leads to a significantly faster dynamics. 

Overall, this analysis through trajectories highlights an interesting point: nonadiabatic effects in the vicinity of a conical intersection induce what we can refer to as ``partial destructive interference'', manifesting itself with a depletion of nuclear density/trajectories, and, at the same time, this effect counteracts the formation of a zero-density line that would be the signature of a topological phase.

\section*{acknowledgement}
This work was supported by the ANR Q-DeLight project, Grant No. ANR-20-CE29-0014 of the French Agence Nationale de la Recherche.

\end{document}